\documentclass[12pt]{article}

\usepackage{amssymb}
\usepackage{amsmath}
\usepackage{latexsym}
\usepackage{yfonts}

\catcode`@=11
\renewcommand{\section}{\@startsection{section}{1}{0pt}{\medskipamount}
{\medskipamount}{\large\bf}} \numberwithin{equation}{section}
\catcode`@=12

\def\beq{\begin{eqnarray}}    
\def\eeq{\end{eqnarray}}      

\def\sDet{\,\mbox{sDet}\,}              

\def\={\ =\ }

\begin{document}

\begin{center}

{\Large\bf Superfield Hamiltonian  quantization in terms of quantum antibrackets}

\vspace{8mm}

{\bf (In respectful memory of Professor Raymond Stora)}

\vspace{18mm}

{\large Igor A. Batalin$^{(a,b)}\footnote{E-mail:
batalin@lpi.ru}$\; and \;
Peter M. Lavrov$^{(b, c)}\footnote{E-mail:
lavrov@tspu.edu.ru}$\;
}

\vspace{8mm}

\noindent ${{}^{(a)}}$
{\em P.N. Lebedev Physical Institute,\\
Leninsky Prospect \ 53, 119 991 Moscow, Russia}

\noindent  ${{}^{(b)}}
${\em
Tomsk State Pedagogical University,\\
Kievskaya St.\ 60, 634061 Tomsk, Russia}

\noindent  ${{}^{(c)}}
${\em
National Research Tomsk State  University,\\
Lenin Av.\ 36, 634050 Tomsk, Russia}

\vspace{20mm}

\begin{abstract}
\noindent We develop a new version of the superfield Hamiltonian
quantization. The main new feature is that the BRST-BFV charge and
the gauge fixing Fermion are introduced on equal footing within the
sigma model approach, which provides for the actual use of the
quantum/derived antibrackets.  We study in detail  the generating
equations for the quantum antibrackets and their primed
counterparts. We discuss  the finite quantum anticanonical
transformations generated by the quantum antibracket.
\end{abstract}

\end{center}

\vfill

\noindent {\sl Keywords:} Hamiltonian quantization, superfield,
quantum antibracket
\\

\noindent PACS numbers: 11.10.Ef, 11.15.Bt
\newpage

\section{Introduction}

In the present paper we develop further the Hamiltonian superfield
quantization suggested in \cite{BBD,BBD1,GD}. The main new feature is
that the BRST-BFV charge and the gauge fixing fermion are introduced
on equal footing within the sigma model approach
\cite{AKSZ,CF,BM,BM4}, which provides for the actual use of the
quantum/derived antibrackets.  We study in detail the generating
equations  for the quantum antibrackets and their primed
counterparts, as well as the finite quantum anticanonical
transformations generated by the quantum antibracket. In this
connection we note also that the quantum antibrackets yield  the
effective means of representation of gauge fields via suitable BRST
operator in the approach recently proposed \cite{BL}.

As the BRST-BFV supersymmetry plays its fundamental role in the
generalized Hamiltonian formalism for dynamical systems with
first-class constraints \cite{BF}, it appears as a quite natural
idea to use the BRST-BFV superfields as to develop the
sigma-model-like approach specific to the topological field
theories. With these regards, we suggest a total superfield action
as a sum of the two dual sigma models related to the BRST-BFV charge
and to the gauge fixing Fermion, respectively.  In this way, we
reproduce the superfield covariant derivative in kinetic part of the
Hamiltonian action, while the unitarizing Hamiltonian itself appears
in the specific  form of a supercommutator of the BRST -BFV charge
and the gauge fixing Fermion. The latter general form of the
unitarizing Hamiltonian is a characteristic  feature of the
generalized Hamiltonian formalism in its form invariant under
reparametrizations of time. Then, the main observation is that the
mentioned general form of the unitarizing Hamiltonian rewrites  in a
natural way entirely in terms of the two dual quantum antibrackets
known  to mathematicians  as "derived brackets".   That  is a
"synthetic" object constructed of double supercommutators  of  its
entries and the generating nilpotent Fermion. These derived brackets
have been introduced by mathematicians in \cite{K-S1,K-S2} (for
further discussion see also \cite{Vo,CSch,Bering}), and then,
independently, by physicists in \cite{BM1,BM2,BM5}. These objects
have very nice algebraic properties such as the generalized Jacobi
relations and the modified Leibnitz rule. In terms of the dual
quantum antibrackets,  the  unitarizing  Hamiltonian splits
additively into two commuting parts, which implies the respective
multiplicative splitting  as to the evolution operator. We study in
detail the generating equations  for all the quantum antibrackets
and their primed counterparts  as well.
\\

\section{Classical action for dual sigma models}

Let
\beq
\label{s1}
z^{ A }, \quad \varepsilon( z^{ A } ) : =  \varepsilon_{ A }, \quad A = 1, ... , 2N, 
\eeq
be a set of canonical pairs coordinate-momentum of a dynamical system with
constraints.

Let us consider the two dual partial actions of the respective sigma
models for  superfields $z^{ A }( t, \theta )$,
\footnote{Here and in what follows we use the following normalization of Berezin's integral
$\int d\theta \;\!\theta=1$.}
\beq
\label{s2}
&&\Sigma_{\Omega}  :=  \int  dt  d\theta  [ V_{ A }  \theta
\partial_{ t }
z^{ A }  ( -1 )^{ \varepsilon_{ A } }   +  \Omega  ], \\  
\label{s3}
&&\Sigma_{\Psi}  :=  \int  dt  d\theta  [V_{ A }  \partial_{ \theta }  z^{
A }  ( -1 )^{ \varepsilon_{ A } }  -  \Psi].      
\eeq
Here in (\ref{s2}), (\ref{s3}) $t$ and $\theta$ is a Boson and Fermion  time variable,
\beq
\label{s4}
\varepsilon( t )  =  0, \quad  \varepsilon (\theta)   =  1,     
\eeq
respectively;
\beq
\label{s5}
V_{ A } = V_{ A }( z ),  \quad  \varepsilon( V_{ A } )  : =   \varepsilon_{ A } , 
\eeq
is  a  symplectic  potential  whose shifted  vorticity  determines the
respective covariant  symplectic metric
\beq
\label{s6}
\omega_{ AB }  :=  \partial_{ A } V_{ B }  +  \partial_{ B } V_{ A} ( -1 )^{
( \varepsilon_{ A } + 1 ) ( \varepsilon_{ B } + 1 ) }   ;    
\eeq $\Omega$ and $\Psi$ is a BRST-BFV charge and a gauge-fixing
Fermion, respectively,
\beq
\label{s7}
&&\varepsilon( \Omega ) = 1,  \quad \{ \Omega, \Omega  \} = 0,  \\  
\label{s8}
&&\varepsilon( \Psi )  =  1,  \quad  \{ \Psi,  \Psi  \} = 0,    
\eeq
where a Poisson bracket is defined as usual with the contravariant
symplectic  metric $\omega^{ AB }$ inverse to (\ref{s6}),
\beq
\label{s9}
\{  F,  G  \}  :=  F \;\overleftarrow{ \partial_{ A }}\;\! \omega^{ AB }
\overrightarrow{\partial_{ B }} \; G.   
\eeq
Now, the complete superfield action is defined as a sum of the partial
actions (\ref{s2}) and (\ref{s3}),
\beq
\label{s10}
\Sigma  :=  \Sigma_{ \Omega }  +  \Sigma_{ \Psi }  =  \int  dt  d\theta [
V_{A} D z^{ A } ( -1 )^{ \varepsilon_{A} }  +  Q ],     
\eeq
where
\beq
\label{s11}
D  :=  \partial_{ \theta }  +  \theta \partial_{ t }, \quad  \varepsilon( D ) = 1,
\quad D^{ 2 }  =  \frac{ 1 }{ 2 } [ D, D ]  =  \partial_{ t },       
\eeq
is a superfield covariant derivative,   and  a total
Fermion generator ("Hamiltonian"),
\beq
\label{s12}
Q  :=  \Omega  -  \Psi  , \quad  \varepsilon( Q )  =  1,  \quad
\{ Q,  Q \} = -2 \{ \Omega,\Psi  \},      
\eeq
respectively.  The complete action (\ref{s10}) yields the following superfield
equation of motion
\beq
\label{s13}
D z^{ A }  -  \{ Q, z^{A} \}  =  0.     
\eeq
Due to (\ref{s11}), (\ref{s12}), it follows immediately from (\ref{s13})
\beq
\label{s14}
\partial_{ t } z^{ A } + \{  \mathcal{ H },  z^{ A } \}  =  0,    
\eeq
where
\beq
\label{s15}
\mathcal{ H }  := - \frac{ 1 }{ 2 } \{ Q, Q \}  =  \{  \Omega,
\Psi  \},    
\eeq
is a complete Boson Hamiltonian. The Hamiltonian (\ref{s15}) provides for correct
dynamical description
for general system with first-class constraints, being  the time sector
$(q^{ 0 }, p_{ 0 } )$ included into the original set (\ref{s1}) as well,  so that
the time  first-class constraint has the form
\beq
\label{s16}
T_{ 0 } :=  p_{ 0 }  +  H_{ 0 },     
\eeq
with $H_{ 0 }( z )$ being an original Hamiltonian \cite{M,M1}.

Now, let us consider in short the supersymmetry properties of the
superfield action (\ref{s10}). It appears as a natural idea to
consider the "BRST" variation of the superfield $z^{A}( t, \theta )$
in the form of a $\theta$-translation
\beq
\label{s17}
\delta z^{A}(t, \theta) : =  z^{A} \overleftarrow{ \partial_{ \theta }}\;\! \mu\;\!.
\eeq
As far as a Fermion parameter $\mu$ is  a constant,  the Jacobian of
the transformation (\ref{s17})
equals to one,  because the derivative of the delta function of
$\theta - \theta'$ equals to one.
On the other hand, by choosing $\mu$ in the form
\beq
\label{s18}
\mu   :=   \frac{ i }{ \hbar }  \int dt d\theta \;\theta \;\delta Q( z ( t,
\theta ) )\;\!,
\eeq
with $\delta Q$ being a "desired" functional variation of $Q$ (do not
confuse with (\ref{s17})\;\!!),  we
find the Jacobian
\beq
\label{s19} J  =  1  +  \frac{ i }{ \hbar }  \int dt d\theta
\;\theta \;( z^{ A } \overleftarrow{\partial_{\theta}} ) ( \delta Q
( z(t, \theta ) )\;\!\overleftarrow{\partial_A} ) ( -1 )^{
\varepsilon_{ A } }  =
1 - \frac{i}{\hbar} \int dt d\theta \;\delta Q( z( t, \theta))\;\!,
\eeq
which reproduces the "desired" variation of the superfield action
(\ref{s10}). Here in (\ref{s19}),  we
have integrated by part over $\theta$
\beq
\nonumber
&&\int  d\theta\; \theta \;(  z^{ A } \overleftarrow{\partial_{ \theta }}  )  (
\delta Q \overleftarrow{\partial_A} ) ( -1 )^{ \varepsilon_{ A } } =
-  \int d\theta \;\theta \;(  \delta Q \overleftarrow{\partial_A }  )  (  z^{
A } \overleftarrow{\partial_{ \theta }} )  =\\
\label{s20} &&\qquad\qquad = -  \int  d\theta \;\theta \;( \delta Q
\overleftarrow{\partial_{\theta }}  )  = -  \int  d\theta\; \delta
Q\;\!. \eeq As to the variation  of kinetic part  of the superfield
action (\ref{s10}) under the variation  (\ref{s17}) with any $\mu$,
it has the form ( $\omega_{ AB }  =  {\rm const}( z )$ )
\beq
\label{s21} -  \mu \int dt d\theta \; (\partial_{\theta } z^{ B } )
\; \omega_{ BA } \;\! D z^{ A }  ( -1 )^{ \varepsilon_{ A } }  = -
\mu \int dt d\theta  \;(\partial_{\theta } z^{ B } )  \big( \omega_{
BA} D z^{ A }  ( -1 )^{ \varepsilon_{ A } }  +  \partial_{ B } Q
\big).
\eeq
where we have taken into account  that
\beq \label{s22}
\int d\theta \;( \partial_{ \theta } z^{ B } ) \partial_{ B } Q  =
\int d\theta \;\partial_{\theta } Q  =  0.
\eeq
The expression in
the second parentheses in  the right-hand side in (\ref{s21}) is
nothing else but the left-hand side of the superfield equations of
motion for $ z^{ B } ( t, \theta )$. Thus we have shown that the
kinetic part of the superfield action (\ref{s10}) is invariant under
the variation (\ref{s17}) with any $\mu$, at the extremals of the
whole superfield action (\ref{s10}).  On the other hand,  the
variation of the rest of the superfield action (\ref{s10}) under the
variation (\ref{s17}) with any $\mu$ vanishes trivially due to
(\ref{s22}),  while  the $\mu$ (\ref{s18}) yields the Jacobian
(\ref{s19}) that reproduces exactly the "desired" functional
variation of the superfield  action (\ref{s10}) under the "desired "
variation $\delta Q$.  As usual in the theories with  weak  global
supersymmetries,  the statements proven are sufficient to conclude
that the superfield path integral constructed of the superfield
action (\ref{s10}) is independent of the "desired" variations of the
gauge fixing Fermion $\Psi$.  The latter  conclusion can be
confirmed in an independent way via the well-known proof within the
component formalism (see Appendix).
\\

\section{Quantum dynamics and quantum antibrackets}

As to the quantum description, we proceed with the operator valued form of
the equations of motion (\ref{s13}), (\ref{s14}), and the Hamiltonian (\ref{s15}),
where all Poisson brackets
should be replaced by the respective
commutators multiplied by $( i \hbar )^{-1}$,
\beq
\label{s2.1}
\{  F,  G  \} \; \rightarrow \; ( i \hbar )^{-1}  [ F,  G ],  \quad
  [ F, G ]  :=  F G  -  G F( -1 )^{ \varepsilon_{ F } \varepsilon_{ G } }.      
\eeq
For the sake of simplicity, we assume in what follows below that
\beq
\label{s2.2}
( i \hbar )^{ -1 }  [ z^{ A } ( t ),  z^{ B }( t ) ]  =  \omega^{ AB }   =
{\rm const}( z ).        
\eeq
Due to (\ref{s2.1}), the Hamiltonian (\ref{s15}) takes the operator valued form
\beq
\label{s2.3}
\mathcal{ H }  :=   ( i \hbar )^{ -1 }  [ \Omega,  \Psi ]. 
\eeq
Then the equation of motion for an operator $A( z )$ reads
\beq
\label{s2.4}
\partial_{ t } A  =  ( i \hbar )^{ -2 }  [ A, [ \Omega, \Psi
] ].     
\eeq
For any two operators F and G, let us define their quantum
$\Omega$-antibracket \cite{BM1}
\beq
\nonumber
( F, G )_{ \Omega }  :&=&  \frac{ 1 }{ 2 } (  [ F, [ \Omega, G ] ] -
( F \leftrightarrow G )  ( -1 )^{ ( \varepsilon_{ F } + 1 ) ( \varepsilon_{ G } + 1
) }  )  =\\
\label{s2.5}
&=&
 [ [ \Omega , F ], G ] ( -1 )^{ \varepsilon_{ F } + 1 }  +
 \frac{ 1 }{ 2} [ \Omega, [ F, G ] ]  ( -1 )^{ \varepsilon_{ F } }.     
\eeq
By definition (\ref{s2.5}) it follows an important formula
\beq
\label{s2.6}
[ \Omega,  ( F,  G )_{ \Omega } ]  =  [ [\Omega,  F], [ \Omega, G ] ].  
\eeq
These quantum antibrackets have very nice algebraic properties. First of
all, we mention their Jacobi identity in a purely Boson sector
\beq
\label{s2.7}
6 ( B, ( B, B )_{ \Omega } )_{ \Omega }  =  [ [ B, (B, B)_{ \Omega } ],
\Omega ] , \quad  \varepsilon( B )  =  0.    
\eeq
Then we apply the differential polarization procedure. By choosing the Boson
B in the form
\beq
\label{s2.8}
B  =  \alpha  F +  \beta  G  +  \gamma  H ,     
\eeq
where $F, G, H$  are  any operators whose Grassmann parities coincide with the
ones of the parameters
$\alpha, \beta, \gamma$, respectively, we  then apply the operator
\beq
\label{s2.9}
\partial_{ \alpha } \partial_{ \beta } \partial_{ \gamma }
( -1 )^{  ( \varepsilon_{ \alpha }  +  1 ) ( \varepsilon_{ \gamma }  +  1 )
+  \varepsilon_{ \beta }  +  1  },  
\eeq
to the relation (\ref{s2.7}),  to  derive the general form of the Jacobi identity
\beq
\label{s2.10}
( F,  ( G,  H )_{ \Omega } )_{ \Omega }  ( -1 )^{ ( \varepsilon_{ F } + 1 )
( \varepsilon_{ H } + 1 ) }  +  {\rm cycle}( F, G, H )  =
 -  \frac{ 1 }{ 2 }  [  ( F, G, H )_{ \Omega }  ( -1 )^{ ( \varepsilon_{ F
} + 1 ) ( \varepsilon_{ H } + 1) } ,  \Omega  ],  
\eeq
where
\beq
\label{s2.11}
( F, G, H )_{ \Omega }  =  \frac{ 1 }{ 3 }  ( -1 )^{( \varepsilon_{ F } +
1 ) ( \varepsilon_{ H } + 1 )}
([ F,  ( G,  H )_{ \Omega } ]  ( -1 )^{ \varepsilon_{ G } + \varepsilon_{
F } ( \varepsilon_{ H } + 1 ) }  +  {\rm cycle}( F, G, H )),  
\eeq
is the so-called quantum 3 - antibracket.
The generalized Leibnitz rule for the quantum antibracket  reads
\beq
\nonumber
\label{s2.12}
&&( FG, H )_{ \Omega }  -  F ( G, H )_{ \Omega }  -  ( F, H )_{ \Omega } G (
-1 )^{ \varepsilon_{ G } ( \varepsilon_{ H } + 1 ) }  =\\
&&=
\frac{ 1 }{ 2 }  \left(  [ F, H ] [ G, \Omega ]  ( -1 )^{ \varepsilon_{ H }
( \varepsilon_{ G } + 1 ) }  +
[ F, \Omega ] [ G, H ] ( -1 )^{ \varepsilon_{ G } }  \right).   
\eeq
Now, let us turn again to the double commutator in the right-hand side in (\ref{s2.4}).  It
is a remarkable fact that \cite{BM2,BM5}
\beq
\label{s2.13}
[ A, [ \Omega, \Psi ] ]  =  \frac{ 2 }{ 3 } (  ( A, \Psi )_{ \Omega } + ( A,
\Omega )_{ \Psi }  ),     
\eeq
together with\footnote{If one does not assume the nilpotency (\ref{s8})
as for the gauge Fermion $\Psi$,
then the right-hand side of the formula (\ref{s2.14}) becomes
$-   \frac{ 1 }{ 4 }  [  \Omega,  (  \frac{ 1 }{ 2 }  [  \Psi,  \Psi   ],   A
)_{ \Omega  }   ]   =
 -  \frac{ 1 }{ 4 }  {\rm ad}(\Omega )  {\rm ad}_{ \Omega }(  \frac{ 1 }{ 2 }  [
\Psi ,  \Psi   ]  )  A $,
while the  formula (\ref{s2.13}) remains the same.}
\beq
\label{s2.14}
[ {\rm ad}_{ \Omega }( \Psi ),  {\rm ad}_{ \Psi }( \Omega ) ] A  :=  ( \Psi,  ( \Omega,
A )_{ \Psi }  )_{  \Omega }  -  ( \Omega  \leftrightarrow \Psi  )  =  0,   
\eeq
which means in turn that the evolution operator
\beq
\label{s2.15}
\exp \{ - (
i \hbar )^{ - 1 } t \mathcal{ H } \},       
\eeq
does factorize into a product of the two commuting operators defined by the
respective two  terms in the right-hand side in (\ref{s2.13}),
 that is
\beq
 \label{s2.16}
 \exp\left\{ - ( i
\hbar )^{ -2 } \;\!t \;\! \frac{ 2 }{ 3 } \;\!{\rm ad}_{ \Omega }( \Psi ) \right\},  
\eeq
and
\beq
\label{s2.17}
\exp\left\{ - ( i\hbar )^{ -2 } \;\! t\;\!
\frac{ 2 }{ 3 }\;\! {\rm ad}_{ \Psi }( \Omega ) \right\},   
\eeq
 each of which being  a quantum anticanonical  transformation generated by a
quantum antibracket.

It follows from the first equation in (\ref{s12}) that the
operator $Q$ does satisfy the closed equation
\beq
\label{n1}
({\cal G}  , {\cal G}  )_{ Q }  = - ( i \hbar )^{2}  Q, 
\eeq
where the $Q$ - quantum antibracket, $( F, G )_{ Q }$, is defined in
(\ref{s2.5}) with $Q$ standing for $\Omega$, while the ${\cal G}$ in the
left-hand side in (\ref{n1}) is the total ghost number operator
\beq
\label{n2}
[{\cal G},  \Omega ]  =  i \hbar  \Omega,  \quad
[{\cal G} ,  \Psi ]  =  - i \hbar  \Psi.    
\eeq

\section{Generating operator for higher quantum
antibrackets}

Let us consider a chain of operators $f_{ a }$ and parameters $\lambda^{ a }$,
\beq
\label{s3.1}
\{  f_{ a }( z );  \lambda^{ a } | \varepsilon( F_{ a } ) = \varepsilon(
\lambda ^{ a } ) :=  \varepsilon_{ a },\;\;  a  =  1 , ... , n , ...   \}, 
\eeq to define the Fermion  nilpotent  generating operator
\beq
\label{s3.2} \Omega( \lambda )  : =  \exp\{ - F \}  \Omega
\exp\{ F \},  \quad    F  :=  f_{ a }\lambda^{ a },   
\eeq
\beq
\label{s3.3}
\varepsilon( \Omega ( \lambda) )  :=  1,   \quad   [  \Omega( \lambda ),  \Omega(
\lambda ) ]  =  0 .    
\eeq
In terms of the generating operator (\ref{s3.2}) the $n-th$ quantum antibracket  is
defined as \cite{BM2}
\beq
\label{s3.4}
(  f_{ a_{ 1 } }, ... , f_{ a_{ n } }  )_{ \Omega }  :=
-  \Omega( \lambda)
\overleftarrow{\partial }_{ a_{ 1 } }  ...  \overleftarrow{\partial }_{ a_{ n } }
|_{ \lambda = 0 }  ( -1 )^{ E_{ n } },    
\eeq
where
\beq
\label{s3.5}
E_{ n }  :=  \sum_{ k = 0 }^{ [ \frac{ n - 1 }{ 2 }  ] } \varepsilon_{ a_{ 2
k + 1 } }.     
\eeq
In more detail, we have
\beq
\label{s3.6}
(  f_{ a_{ 1 } }, ... , f_{ a_{ n } }  )_{ \Omega }  =  -  [  ...  [ \Omega,
f_{ b_{ 1 } } ],  ... , f_{ b_{ n } } ]
 S^{ b_{ n}  ...  b_{ 1 } }_{ a_{ 1 }  ...  a_{ n } }  ( -1 )^{ E_{ n } }, 
\eeq
with the symmetrizer being defined as
\beq
\label{s3.7}
n!  S^{ b_{ n }  ...  b_{ 1 } } _{ a_{ 1 }  ...  a_{ n } }  :=   (
\lambda^{ b_{ n } }  ...  \lambda^{ b_{ 1 } }
\overleftarrow{\partial }_{ a_{ 1 } }  ... \overleftarrow{\partial }_{ a_{ n } }
),  \quad       \partial_{ a }  :=  \frac{ \partial }{ \partial \lambda^{ a } }. 
\eeq
All the Jacobi relations for higher quantum antibrackets are accumulated in
the nilpotency equation (\ref{s3.3}).
In terms of the generating operator (\ref{s3.2}) together with the operator
\beq
\label{s3.8}
R_{ a }( \lambda)  :=
\exp\{ - F \} \left(  \exp\{ F \} \overleftarrow{\partial }_{ a }\right),      
\eeq
the following closed set of  the generating  equations holds
\beq
\label{s3.9}
\Omega( \lambda ) \overleftarrow{\partial }_{ a }   =   [ \Omega( \lambda ),
R_{ a }( \lambda ) ],  
\eeq
\beq
\label{s3.10}
R_{ a }( \lambda ) \overleftarrow{\partial }_{ b}  -  ( a \leftrightarrow b ) ( -1 )^{
\varepsilon_{ a } \varepsilon_{ b } }  =
[ R_{ a }( \lambda ),  R_{ b }( \lambda ) ] ,  
\eeq
\beq
\label{s3.11}
\Omega( \lambda  =  0 )  =  \Omega,  \quad  R_{ a }( \lambda  =  0 )  =   f_{ a
}.    
\eeq
In turn, these generating equations  do imply  further equations for primed
quantum antibrackets defined as
\beq
\label{s3.12}
( f_{ a _{1} },  ...  , f_{ a_{ n } }  )'_{ \Omega }  :=  -  \Omega( \lambda
)
\overleftarrow{\partial }_{ a_{ 1 } } ... \overleftarrow{\partial }_{ a_{ n } }  (
-1 )^{ E_{ n } }.   
\eeq
In particular, for primed quantum 2-antibracket we get
\beq
\label{s3.12}
( f_{ a },  f_{ b } )'_{ \Omega }   =   (  R_{ a }( \lambda ),  R_{ b }(
\lambda )  )_{ \Omega( \lambda ) }  -
 \frac{ 1 }{ 2 } [  \Omega( \lambda ),  R_{ a }( \lambda ) \overleftarrow{
\partial }_{ b }  +
( a  \leftrightarrow b )  ( -1 )^{ \varepsilon_{ a } \varepsilon_{ b } }  ]  ( -1
)^{ \varepsilon_{ a } }.    
\eeq

\section{Finite quantum anticanonical transformations}

In Section 3, we have already mentioned finite quantum anticanonical
transformations (\ref{s2.16}), (\ref{s2.17}).  Now we are in a position to present such
transformations explicitly in their most general setting.  Let $\lambda$ be
a  boson parameter,  $\varepsilon( \lambda ) = 0$.  Given an operator $A$,
define then the  transformed operator  as
\beq
\label{s5.1}
A'  :=  \exp\{ \lambda \; {\rm ad}_{ \Omega }( \Psi ) \}  A,    
\eeq
to satisfy the equation
\beq
\label{s5.2}
\partial_{ \lambda }  A'  =  ( \Psi,  A' )_{ \Omega },  \quad  A' ( \lambda  =  0
)  =  A.   
\eeq
Its explicit  solution has the form \cite{BT}
\beq
&&A'  ={\tilde A}(\lambda)
\label{s5.3}
-
\frac{ 1 }{ 2 }  \int_{ 0 }^{ \lambda }  d\lambda'  \exp\left\{  \frac{ \lambda
-  \lambda' }{ 2 }  [ \Omega, \Psi  ]  \right\}
[  \Omega,   [ \Psi,  {\tilde A}(\lambda')  ]  ]
\exp\left\{ - \frac{ \lambda  -  \lambda' }{ 2 }  [ \Omega, \Psi  ]  \right\},  
\eeq
where
\beq
\label{s5.0}
{\tilde A}(\lambda)=\exp\{ \lambda [ \Omega, \Psi  ] \}  A   \exp\{ -
\lambda  [ \Omega, \Psi  ]  \}.
\eeq
By interchanging $\Omega \leftrightarrow \Psi$ in (\ref{s5.3}),
we get the transformation dual to (\ref{s5.3}).
In this way, the operators (\ref{s2.16}), (\ref{s2.17}) are reproduced at
\beq
\label{s5.4}
\lambda =  -  ( i \hbar )^{ -2 }  t  \frac{ 2 }{ 3}.       
\eeq

Also, notice that the quantum antibracket of the two transformed
operators $A'$ and $B'$ satisfies the equation \cite{BT}
\beq
\label{s5.5}
\partial_{ \lambda }  ( A',  B' )_{ \Omega }  =  (  \Psi,  ( A',  B' )_{
\Omega }  )_{ \Omega }  +
 \frac{ 1 }{ 2 }  [  ( \Psi,  A',  B' )_{ \Omega },  \Omega  ]\;,      
\eeq
that follows from (\ref{s5.2}) for $A'$ and $B'$, together with the Jacobi
relation (\ref{s2.10}).

Explicit solution to the equation (\ref{s5.5})  has the form
\beq
\label{s5.7}
(  A'( \lambda ),  B'( \lambda )  )_{ \Omega }   =
( A,  B )'_{ \Omega} (\lambda )  +
 \frac{ 1 }{ 2 }  \int _{ 0 }^{ \lambda }  d\lambda'  \exp\{  ( \lambda  -
\lambda' )  {\rm ad}_{ \Omega }( \Psi )  \}
[(  \Psi,  A'( \lambda' ),  B' ( \lambda' )  )_{ \Omega },\Omega],     
\eeq
where all primed operators are defined similarly to (\ref{s5.1}).
For instance, the first term in the  right-hand side   in (\ref{s5.7}) is decoded as
\beq
\label{s5.8}
\exp\{ \lambda \;{\rm ad}_{ \Omega }( \Psi ) \} (  A  , B )_{ \Omega },   
\eeq
not to be confused with (\ref{s3.12})!
\\

\section{Finite transformations of general open group:
integrating arbitrary involutions [15, 21]}

Let
\beq
\label{s6.1}
\{  \phi_{ a } |  \varepsilon( \phi_{ a } )  :=
\varepsilon_{ a }  = \varepsilon( T_{ a } ) \},      
\eeq
be a set of parameters of gauge transformations generated by the
first-class constraints
$T_{a}$  encoded in the  BRST-BFV operator $\Omega$.  Let us consider
the general  Lie equation
for an operator valued transformation  $A_{ 0 }\;\rightarrow \; A( \phi )$,
\beq
\label{s6.2}
&&A( \phi ) \overleftarrow{\partial }_{ a }  =  ( i \hbar )^{ -1 }  [
A( \phi ),  Y_{ a }( \phi )  ], \\    
\label{s6.3}
&& A( \phi   =  0  )  =   A_{ 0 }, \\  
\label{s6.4}
&&\partial_{ a }  :=  \frac{ \partial }{ \partial \phi^{ a } } ,\quad
\varepsilon( Y_{ a } )  := \varepsilon_{ a }.  
\eeq
Integrability of that equation requires
\beq
\label{s6.5}
Y_{ a } \overleftarrow{\partial }_{ b }  -  ( a \leftrightarrow b )  ( -1 )^{
\varepsilon_{ a } \varepsilon_{ b }  }  =
( i \hbar )^{ -1 }  [ Y_{ a },  Y_{ b } ].        
\eeq
We choose the operators $Y_{ a }$ in the form generated by the one
$\Omega$,
\beq
\label{s6.6}
&&Y_{ a }( \phi )  :=  ( i \hbar )^{ -1 }  [ \Omega,  \Omega_{ a }(
\phi ) ],\\ 
\label{s6.7}
&& \varepsilon( \Omega_{ a } )  :=  \varepsilon _{ a }  + 1.    
\eeq
The form (\ref{s6.6}) implies that
\beq
\label{s6.8}
[ \Omega,   A_{ 0 } ]  =  0  \quad\Rightarrow\quad [ \Omega,   A( \phi ) ]  =  0. 
\eeq
Then,  the integrability (\ref{s6.5}) together with the choice (\ref{s6.6})
implies that
\beq
\label{s6.9}
A \overleftarrow{\partial }_{ a }  =  ( i \hbar )^{ -2 }  (  A ,
\Omega_{ a } )_{ \Omega }  +
( i \hbar )^{ -2 }  \frac{ 1 }{ 2 }  [  [  A,  \Omega_{ a }  ],
\Omega ]  ( -1 )^{  \varepsilon_{ a }  },    
\eeq
\beq
\label{s6.10}
\Omega_{ a } \overleftarrow{ \partial }_{ b }  -  ( a \leftrightarrow b )  ( -1
)^{ \varepsilon_{ a } \varepsilon_{ b }  }  =
( i \hbar )^{ -2 } ( \Omega_{ a },  \Omega_{ b } )_{ \Omega }  -
 \frac{ 1 }{ 2 } ( i \hbar )^{ -1 } [ \Omega,  \Omega_{ ab} ], 
\eeq
In its own turn,  the  integrability condition (\ref{s6.10}) requires
further integrability conditions, and so on.
It is a remarkable fact that all these subsequent integrability
conditions are naturally accumulated in a single quantum master equation
\beq
\label{s6.11}
(  S,  S  )_{ \Delta  }   =   i  \hbar  [ \Delta,  S ],\quad
\varepsilon( S ) =  0,   
\eeq
where
\beq
\label{s6.12}
\Delta   :=   \Omega   +   \eta^{ a } \pi_{ a }  ( -1 )^{
\varepsilon_{ a } }, \quad    \Delta^{ 2 }  =  0,    
\eeq
with
\beq
\label{s6.13}
\pi_{a},  \quad  \varepsilon( \pi_{ a } )   :=   \varepsilon_{ a }, 
\eeq
being  momenta  canonically conjugated to  $\phi^{ a }$,
\beq
\label{s6.14}
[  \phi^{ a },  \pi_{ b }  ]   =   i  \hbar  \delta^{ a }_{ b }, 
\eeq
and
\beq
\label{s6.15}
\eta^{ a },  \quad  \varepsilon( \eta^{ a } )   :=  \varepsilon_{ a }  + 1, 
\eeq
being  new ghost variables viewed as parameters.  As we have by
definition
\beq
\label{s6.16}
(  S,  S  )_{ \Delta }   =   [  S,  [  \Delta,  S  ]  ],  
\eeq
the master equation (\ref{s6.11}) implies
\beq
\label{s6.17}
[  \Delta,  (  S,  S  )_{ \Delta }   ]   =   0  \quad  \Rightarrow\quad  [  \Delta,  S
]^{ 2 }  =  0.     
\eeq

Now,  let  ${\cal G}$ be the standard ghost number operator,
\beq
\label{s6.18}
[ {\cal G},  \Omega  ]   =   i \hbar  \Omega.    
\eeq
Let us seek for a solution to the  master equation (\ref{s6.11}) in the
form of an $\eta$ -power series expansion,
\beq
\nonumber
\label{s6.19}
S( \phi,  \eta  )   &=&   {\cal G}   +   \eta^{ a } \;\Omega_{ a }( \phi )   +
\frac{ 1 }{ 2 } \eta^{ b } \eta^{ a } ( -1 )^{ \varepsilon_{ b }\;
} \Omega_{ ab }( \phi )  +
 \frac{ 1 }{ 6 } \eta^{ c } \eta^{ b } \eta^{ a }
( -1 )^{  \varepsilon_{ b }  +  \varepsilon_{ a } \varepsilon_{ c }\;
} \Omega_{ abc }( \phi  )   +   ...   +\\
\label{s6.19}
&&+\frac{ 1 }{ n ! }  \eta^{ a_{ n } }  ...  \eta^{ a_{ 1 } }  ( -1
)^{ \varepsilon_{ n } }
\Omega_{ a_{ 1 } ...  a_{ n } }( \phi  )   +   ...    ,    
\eeq
where
\beq
\label{s6.20}
\varepsilon_{ n }  :=   \sum_{ k  =  1 }^{  [  \frac{ n }{ 2 }  ]  }
\varepsilon_{ a_{ 2 k } }  +
 \sum_{ k  =  1 }^{  [  \frac{ n - 1 }{ 2 }  ]  }   \varepsilon_{
a_{ 2 k - 1 } } \varepsilon_{ a_{ 2 k + 1 } } .    
\eeq
The coefficient operators  $\Omega_{ a_{ 1 }  ...  a_{ n } }$  in the
expansion (\ref{s6.19}) have the properties
\beq
\label{s6.21}
&&\varepsilon(  \Omega _{ a_{ 1 }  ...  a_{ n } }  )   =
\varepsilon_{ a_{ 1 } }  +   ...   +  \varepsilon_{ a_{ n } }  +  n, \\ 
\label{s6.22}
&& [ {\cal G},  \Omega_{ a_{ 1 }  ...   a_{ n } } ]   =   - n  i \hbar\;
\Omega_{ a_{ 1 }  ...  a_{ n } }.    
\eeq
In the zeroth and first orders in $\eta$, the master equation (\ref{s6.11})
is satisfied identically.  However,
in the second order, it yields exactly (\ref{s6.10}).  In the third order
in $\eta$ it yields
\beq
\nonumber
&&\big(   \partial_{ a } \Omega _{ bc }  +  \frac{ 1 }{ 2 }  ( i  \hbar
)^{ -2 } ( \Omega_{ a },  \Omega_{ bc} )_{ \Omega }  -
\frac{ 1 }{ 12 }  ( i \hbar )^{ -2 }  [   [  \Omega_{ ab },
\Omega_{ c } ],  \Omega   ]\big)     ( -1 )^{ \varepsilon_{ a }
\varepsilon_{ c } }  +
{\rm  cycle} ( a, b, c )   = \\
\label{s6.23}
&&=  - ( i \hbar )^{ -3 } ( \Omega_{ a },
\Omega_{ b }, \Omega_{ c } )_{ \Omega }  ( -1 )^{ \varepsilon_{ a }
\varepsilon_{ c } } -
 \frac{ 2 }{ 3 } ( i \hbar )^{ -1 }  [  \Omega_{ abc },  \Omega], 
\eeq
which is exactly the integrability condition to (\ref{s6.10}).
A natural  automorphism   of  the master equation (\ref{s6.11}) is given
by
\beq
\label{s6.24}
S \;\rightarrow\;    S'   : =   \exp\{  - ( i  \hbar )^{ -2 }  [ \Delta,  \Xi
]  \} \;  S\; \exp\{  ( i  \hbar )^{ -2 }  [ \Delta,  \Xi  ]  \},   
\eeq
where $\Xi$  is an arbitrary odd operator.  For infinitesimal
transformation we have
\beq
\label{s6.25}
\delta S  =  ( i \hbar )^{ -2 }  [  S, [  \Delta,  \Xi  ]  ]  =  ( i
\hbar )^{ -2 }  \frac{ 2 }{ 3 }  \big(  ( S,  \Xi  )_{ \Delta  }  +  (
S,  \Delta )_{ \Xi }  \big),  
\eeq
\beq
\label{s6.26}
\delta_{ 21 }  S   :=   [   \delta_{ 2 } ,  \delta_{ 1 }  ]  S   =
( i \hbar )^{ -2 }  [  S,  [  \Delta,  \Xi_{ 21 }  ]  ],   
\eeq
\beq
\label{s6.27}
\Xi_{ 21 }   =   ( i  \hbar )^{ -2 }  (  \Xi_{ 2 },  \Xi _{ 1 }  )_{
\Delta }.      
\eeq
If the transformation (\ref{s6.24}) acts transitively on the set of
solutions to the master equation (\ref{s6.11}),
then the general solution is
\beq
\label{s6.28}
S  =   \exp\{ - ( i  \hbar )^{ -2 }  [ \Delta,  \Xi  ]  \} \; {\cal G} \; \exp\{ ( i
\hbar )^{ - 2 } [ \Delta,  \Xi  ]  \}.    
\eeq

Now, let us consider the transformation
\beq
\label{s6.29}
S( \alpha )   :=  \exp\{  i  \alpha  F / \hbar \} \; S \; \exp\{ - i
\alpha  F / \hbar  \},    
\eeq
\beq
\label{s6.30}
\Delta( \alpha )  :=  \exp\{  i  \alpha  F  / \hbar \} \; \Delta \;  \exp\{ - i
\alpha  F / \hbar  \},   
\eeq
where $\alpha$ is  an even parameter, and $F$ is an arbitrary even operator.
If $S$ and $\Delta$ satisfy
the master equation (\ref{s6.11}), then $S( \alpha )$ and $\Delta( \alpha )$ satisfy
the transformed master equation
\beq
\label{s6.31}
(  S( \alpha ),  S( \alpha )  ) _{ \Delta( \alpha ) }   =   i  \hbar  [
\Delta ( \alpha ),  S( \alpha )  ].    
\eeq
If  $F$ is restricted to satisfy itself the master equation (\ref{s6.11}),  i. e.
\beq
\label{s6.32}
(  F,  F  )_{ \Delta  }   =   i  \hbar  [  \Delta,   F  ],   
\eeq
then
\beq
\label{s6.33}
\Delta'' ( \alpha )  +  \Delta'( \alpha )  =  0,    
\eeq
and $\Delta( \alpha )$ in (\ref{s6.30}) reduces to
\beq
\label{s6.34}
\Delta( \alpha )  =  \Delta   +   ( i  \hbar )^{ -1 }  [  \Delta,   F  ]  (
1  -  \exp \{ - \alpha \}  ) .    
\eeq
For $F  =  S$, in particular, S satisfies the master equation (\ref{s6.11})  with
$\Delta$ replaced by $\Delta( \alpha )$
in (\ref{s6.34}), where  $F$ is replaced by $S$.
\\

\section{Discussion}

The main result of the present consideration is that the dynamical
evolution of an arbitrary dynamical system with first-class
constraints is represented entirely in terms of the two dual quantum
antibrackets related to the two nilpotent Fermion operators,  the
BRST-BFV charge and the gauge-fixing Fermion.  Although in the
standard BRST-BFV scheme there is no need to impose the nilpotency
requirement as to the gauge Fermion, in the sigma model approach
developed above that requirement should be imposed certainly on
equal footing upon both the BRST-BFV charge and the gauge fixing
Fermion.  If  one allows for the gauge-fixing  Fermion operator  to
deviate from being nilpotent,  then the closed character of the
description in terms of the dual quantum antibrackets will be failed
immediately. It should be noticed however that the standard
properties of gauge invariance in the physical sector remain
maintained in the latter case, as well.
\\

\section*{Acknowledgments}
\noindent I. A. Batalin would like  to thank  Robert Marnelius  of Chalmers University ( Ret.),
 Klaus Bering of Masaryk
University and Igor Tyutin of Lebedev Institute for interesting
discussions. The work of I. A. Batalin is supported in part by the
RFBR grants 14-01-00489 and 14-02-01171. The work of P. M. Lavrov is
supported by the Ministry of Education and Science of Russian
Federation, project No 2014/387/122.

\appendix
\vspace{5mm}
\section{Component formalism}

Let $\theta$ be a Fermionic time (the BRST - parameter), and let
\beq
\label{r1}
z^{A}( t, \theta ) := z^{A}_{0}( t ) + \theta z^{A}_{1}( t ),  \quad \varepsilon(
z^{A}_{0} ) := \varepsilon_{A}, \quad \varepsilon( z^{A}_{1} ) := \varepsilon_{A}
+ 1,   
\eeq be a component expansion to the superfield in the left-hand
side of the first  in (\ref{r1}). First, let us reproduce in terms
of the component expansion (\ref{r1})  the above Jacobian (\ref{s19}).  The
$\theta$ - translation (\ref{s17}) takes the form
\beq
\label{ra2}
\theta \;\rightarrow \; \theta  +  \mu,     
\eeq
where the "parameter" $\mu$ is chosen in the form of the functional
(\ref{s18}),
\beq
\label{ra3}
\mu = \frac{i}{\hbar} \int dt d\theta \;\theta\;\! \delta Q( z( t, \theta )
) = \frac{i}{\hbar} \int dt \;\delta Q( z_{0}( t ) ).   
\eeq
Here, the translation (\ref{ra2}) induces the component variations
\beq
\label{ra4}
\delta z^{A}_{0}( t ) =  \mu  z^{A}_{1}( t ),   \quad   \delta z^{A}_{1}(
t ) = 0.  
\eeq
Due to the choice (\ref{ra3}), the variations (\ref{ra4}) yield the Jacobian
\beq
\nonumber
J &=& 1 + \int dt \;( \delta z^{A}_{0}( t ) ) \frac{ \overleftarrow{ \delta
} }{ \delta z^{A}_{0}( t ) }  (-1)^{ \varepsilon_{A} }  =\\
\nonumber
&=&
1 + \frac{i}{\hbar} \int  dt \; ( \delta Q( z_{0}( t ) ) )\;\!
\overleftarrow{\partial_A} \;\! z^{A}_{1}( t )  (-1)^{ \varepsilon_{A} } =\\
\nonumber
&=&1 - \frac{i}{\hbar} \int  dt  d\theta \;( \delta Q ( z_{0}( t ) ) )
\overleftarrow{\partial }_{A}  \theta  z^{A}_{1}( t ) =\\
\label{ra5}
&=&
1 - \frac{i}{\hbar} \int dt d\theta \;\delta Q( z ( t, \theta ) ) =(\ref{s19}). 
\eeq
Thus, we have reproduced exactly the Jacobian (\ref{s19}) within the
component formalism.

By substituting the component form (\ref{r1}) into the superfield
action (\ref{s10}), we get
\beq
\label{r2}
\Sigma = \int dt \left[
\frac{1}{2} z^{B}_{0} \omega_{BA} \partial_{t} z^{A}_{0}
+\frac{1}{2} z^{B}_{1} \omega_{BA} z^{A}_{1} (-1)^{ \varepsilon_{A}
} +
z^{B}_{1} \partial_{B} Q( z_{0} )\right],  \quad \omega_{BA} = {\rm const}( z_{0}, z_{1} ). 
\eeq Under the variation in the first in (\ref{ra4}), the variation
of kinetic part  of  the action $\Sigma$ (\ref{r2}) reads \beq
\label{r2.1}
\int dt\;  \mu  z ^{B}_{1}  (  \omega_{BA} \partial_{ t
} z^{A}_{0} + \partial_{B}
z^{A}_{1} \partial_{A} Q( z_{0})  ).     
\eeq
Here in (\ref{r2.1}), the second term does not contribute actually due to the
nilpotency property
\beq
\label{r2.2}
( z^{A}_{1} \partial_{A} )^{2} = 0.     
\eeq
The expression inside the parentheses in (\ref{r2.1}) is nothing else but the
left-hand side of the  motion equation for  $z^{B}_{0}$
as to the action $\Sigma$ (\ref{r2}). Thus, the expression
(\ref{r2.1}) is a component counterpart to (\ref{s21}).

Within the path integral
\beq
\label{r3}
Z =: \int Dz_{0} Dz_{1} \exp\left\{ \frac{i}{\hbar} \Sigma \right\},    
\eeq
by taking the Gaussian integral over $z_{1}$, one arrives at the expression
\beq
\label{r4}
Z = \int Dz_{0}  \sqrt{\sDet( \omega)} \exp\left\{ \frac{i}{\hbar}
\Sigma_{0} \right\},   
\eeq
where the zero-th  sector action $\Sigma_{0}$ is given by
\beq
\label{r5}
\Sigma_{0} := \int dt \left[  \frac{1}{2}  z^{B}_{0} \omega_{BA} \partial_{t}
z^{A}_{0} - \mathcal{H}_{0}  \right],  
\eeq
where the Hamiltonian in the zero-th sector is given by
\beq
\label{r6}
\mathcal{H}_{0} := -  \frac{1}{2} \{ Q, Q \}( z_{0} ).   
\eeq
As usual, the Gaussian integral over $z_{1}$ is equivalent, up to the constant
factor of $ \sqrt{\sDet( \omega)}$, to eliminating the $z_{1}$
component by resolving the classical equation for $z_{1}$,
\beq
\label{r7}
\omega_{BA} z^{A}_{1} (-1)^{ \varepsilon_{A} } + \partial_{B} Q( z_{0} ) =
0,     
\eeq
in the form
\beq
\label{r8}
z^{A}_{1} = - \omega^{AB} \partial_{B} Q( z_{0} ) (-1)^{ \varepsilon_{A} },   
\eeq
which is equivalent to
\beq
\label{r9}
z^{A}_{1} = \{ Q, z^{A}_{0} \}.    
\eeq
By substituting the latter into the component action (\ref{r2}), we get exactly the
zero-th  sector action (\ref{r5}).

Now, let us consider the infinitesimal  BRST-BFV transformation,
\beq
\label{r10}
\delta z^{A}_{0} := \{ z^{A}_{0}, Q ( z_{0} ) \} \mu.     
\eeq
As far as a BRST- BFV parameter $\mu$ is a constant, the action (\ref{r5}) and the
measure in (\ref{r4}) are invariant
under  the transformation  (\ref{r10}).  Then let us choose in (\ref{r10}) the $\mu$ in the
form
\beq
\label{r11}
\mu := \frac{i}{\hbar} \int dt \delta Q( z_{0} ),  
\eeq
where $\delta Q$  is the desired functional variation of $Q$.  As the $\mu$
(\ref{r11})
is not a *function* of the phase
variables $z_{0}( t )$,  the action (\ref{r5}) remains invariant under the
transformation (\ref{r10}), (\ref{r11}).  However  the
functional (\ref{r11}) yields the following infinitesimal  Jacobian to the latter
transformation,
\beq
\label{r12}
J := 1 - \frac{i}{\hbar} \int dt \{ Q,  \delta Q \}( z_{0} ),    
\eeq
which is exactly the desired functional variation of the action (\ref{r5}). Thus,
we have shown that the integral
(\ref{r3}) remains stable under the desired functional variations $\delta Q$.
\\

\end{document}